\documentclass[a4paper, 12pt]{article}

\usepackage{axodraw}
\usepackage[english]{babel}
\usepackage{cite}
\usepackage{a4wide}
\usepackage{amsmath}
\usepackage[charter,cal=cmcal,greekuppercase=italicized]{mathdesign}
\usepackage{dsfont}
\usepackage[T1]{fontenc}
\usepackage{slashed}
\usepackage{setspace}
\usepackage{graphicx}
\usepackage{xcolor}
\usepackage{soul}
\usepackage[final,          
            colorlinks,     
            linkcolor=purple, 
            citecolor=teal, 
            urlcolor=blue  
            ]{hyperref}

\makeatletter
\DeclareRobustCommand*{\bfseries}{%
  \not@math@alphabet\bfseries\mathbf
  \fontseries\bfdefault\selectfont
  \boldmath
}
\makeatother

\newcommand{\Ll}{\ensuremath{\mathrm{L}}}
\newcommand{\Rr}{\ensuremath{\mathrm{R}}}
\newcommand{\tq}{\ensuremath{\mathrm{t}}}
\newcommand{\bq}{\ensuremath{\mathrm{b}}}

\newcommand{\dq}{\ensuremath{\mathrm{d}}}
\newcommand{\uq}{\ensuremath{\mathrm{u}}}

\newcommand{\upsh}[1]{\ensuremath{\mathrm{#1}}}

\newcommand{\SU}{\ensuremath{\mathrm{SU}}}
\newcommand{\U}{\ensuremath{\mathrm{U}}}

\newcommand{\GeV}{\ensuremath{\mathrm{GeV}}}
\newcommand{\TeV}{\ensuremath{\mathrm{TeV}}}

\newcommand{\hc}{\ensuremath{\text{h.\,c.\ }}}

\renewcommand{\Re}{\ensuremath{\operatorname{Re}}}

\newcommand{\vev}{\emph{vev}}
\newcommand{\ie}{i.\,e.~}

\newcommand{\eg}{e.\,g.~}

\usepackage[font=small]{caption}

\begin{document}

\onehalfspacing

\begin{titlepage}

\vspace*{-15mm}
\begin{flushright}
TTP15-033
\end{flushright}
\vspace*{0.7cm}

\begin{center} {
\bfseries\LARGE
Charge and color breaking constraints in the Minimal Supersymmetric
Standard Model associated with the bottom Yukawa coupling}
\\[8mm]
Wolfgang~Gregor~Hollik
\footnote{E-mail: \texttt{wolfgang.hollik@kit.edu}}
\\[1mm]
\end{center}
\vspace*{0.50cm}
\centerline{\itshape
Institut f\"ur Theoretische Teilchenphysik, Karlsruhe Institute of Technology}
\centerline{\itshape
Engesserstra\ss{}e 7, D-76131 Karlsruhe, Germany
\footnote{New address as of October 1st 2015: \itshape Theory Group, DESY Notkestra\ss{}e 85, D-22607 Hamburg}}
\vspace*{1.20cm}

\begin{abstract}
\noindent
Testing the stability of the electroweak vacuum in any extension of
the Standard Model Higgs sector is of great importance to verify the
consistency of the theory. Multi-scalar extensions as the Minimal
Supersymmetric Standard Model generically lead to unstable
configurations in certain regions of parameter space. An exact
minimization of the scalar potential is rather an impossible analytic
task. To give handy analytic constraints, a specific direction in
field space has to be considered which is a simplification that tends
to miss excluded regions, however good to quickly check parameter
points. We describe a yet undescribed class of charge and color
breaking minima as they appear in the Minimal Supersymmetric Standard
Model, exemplarily for the case of non-vanishing bottom squark vacuum
expectation values constraining the combination \(\mu Y_\bq\) in a
non-trivial way. Contrary to famous \(A\)-parameter bounds, we relate
the bottom Yukawa coupling with the supersymmetry breaking
masses. Another bound can be found relating soft breaking masses and
\(\mu\) only. The exclusions follow from the tree-level minimization
and can change dramatically using the one-loop potential. Estimates of
the lifetime of unstable configurations show that they are either
extremely short- or long-lived.
\end{abstract}
\end{titlepage}

\setcounter{footnote}{0}

\section{Introduction}
A complete analysis of the vacuum structure in any quantum field
theory needs a consideration of the effective potential to all orders
which is more than an honorable task. Important contributions to the
effective potential in the Standard Model and supersymmetrized
versions at one and way more loops have been (partially)
determined~\cite{Coleman:1973jx, Jackiw:1974cv, Ford:1992pn,
  Martin:2001vx, Martin:2013gka, Martin:2015eia}. The more loops the
more difficult is also the task to find the global minimum which shall
determine the vacuum state of the theory. Numerical solutions to that
problem exist in the Minimal Supersymmetric Standard Model (MSSM)
where both the effective potential as well as the (expected-to-be)
global minimum are calculated and determined purely
numerically~\cite{Camargo-Molina:2013qva,
  Camargo-Molina:2013sta}. Supersymmetry (SUSY) generically tends to
stabilize the potential as negative fermionic loop contributions are
compensated by the corresponding bosonic ones. The superpartner
spectrum on the other hand brings additional directions in scalar
field space that potentially invalidate the electroweak Higgs vacuum
at the classical level. A physical viable supersymmetric extension has
to take care of the additional parameters in a way that the
``desired'' vacuum is the true vacuum of the theory.

The consideration of the one-loop effective potential, which can be very
efficiently done via the famous formula of Coleman and
Weinberg~\cite{Coleman:1973jx}, leads to a first understanding of
non-trivial minima. We have
\begin{equation}\label{eq:CWpot}
V_\text{CW} = \frac{1}{64\pi^2} \sum_f C_f \operatorname{STr} \left[
\mathcal{M}_f^4 (\phi) \left( \ln\big(\mathcal{M}_f^2(\phi)/Q^2 \big) +
    P_f(\phi) \right) \right],
\end{equation}
where the sum runs over all fields \(f\) in the loop and \(C_f\) counts
gauge degrees of freedom like \(C_\text{quark} = 3\) (spin degrees of
freedom are covered by the supertrace \(\operatorname{STr}\)). The
field-dependent mass eigenvalues \(\mathcal{M}_f(\phi)\) are generically
the eigenvalues of the Hessian matrix of the full scalar potential and
the field \(\phi\) represents any type of scalar field value which is
still present in the masses (do not set remnant field values to zero,
they correspond to vacuum expectation values (\vev{}s) at local or
global minima of the potential). Additionally, there is a polynomial
\(P_f(\phi)\) which is renormalization scheme dependent and in the most
common cases a constant. The renormalization scale is given by \(Q\).

The one-loop potential is known to develop an imaginary
part~\cite{Fujimoto:1982tc, Weinberg:1987vp, Camargo-Molina:2013sta,
  Bobrowski:2014dla} which is of no importance in the discussion of
tunneling times from false to true vacua but opens the access to
non-standard vacua: an imaginary part in the one-loop effective
potential is related to a non-convex tree-level potential at that
point.\footnote{It is actually related to a branch point of the
  logarithm in Eq.~\eqref{eq:CWpot} that appears for a zero mass
  eigenvalue.} A non-convex potential means that the second derivative
is negative which corresponds to a tachyonic mass eigenvalue
\(\mathcal{M}_f^2 (\phi) < 0\). The tachyonic mass, however, would only
be present at the minimum (which by definition is locally convex). So,
the existence of a non-convex direction points towards a minimum in that
direction unless the potential is unbounded from below, which would be
even worse. Finding the critical field value at which the non-convex
direction opens is trivial as we shall see. The question is rather
whether the non-standard minimum is deeper than the standard one and
therefore allows for a vacuum-to-vacuum transition which can be figured
out analytically under certain circumstances.

We first consider the loop corrected Higgs potential in the MSSM
including SUSY loop contributions from the third generation
(s)fermions. The tree-level part is given by the mass terms and the
self-couplings which are gauge couplings. The one-loop part is given by
the logarithms of Eq.~\eqref{eq:CWpot} which also follow from the
direct calculation~\cite{Bobrowski:2014dla}. We borrow the notation
from~\cite{Bobrowski:2014dla} and define the effective potential as
\begin{equation}\label{eq:VeffMSSM}
\begin{aligned}
  V_\text{eff}  =\;& V_0+V_1^{\tilde t}+V_1^{t}+V_1^{\tilde b}+V_1^{b} \\
  =\;& m_{11}^{2\,^\text{tree}}\; |h_\dq^{0}|^2 +
  m_{22}^{2\,^\text{tree}}\; |h_\uq^{0}|^2 - 2 \Re\left(
    m_{12}^{2\,^\text{tree}} \; h_\uq^0 h_\dq^0 \right)
  +\frac{g_1^2+g_2^2}{8} \left( |h_\dq^{0}|^2 - |h_\uq^{0}|^2
  \right)^2 \\
  & + \frac{N_c {\widetilde{M}}_\tq^4}{32\pi^2} \bigg[
  \left(1+x_\tq+y_\tq\right)^2 \ln \left(1+x_\tq+y_\tq\right) +
  \left(1-x_\tq+y_\tq\right)^2 \ln \left(1-x_\tq+y_\tq\right)
  \\
  & \qquad\qquad\; - \left(x_\tq^2 + 2y_\tq\right) \left( 3 - 2
    \ln\left({\widetilde{M}}_\tq^2/Q^2\right) \right) -2 {y_\tq^2}
  \ln\left(y_\tq\right) \;+\;\{\tq \leftrightarrow \bq\} \bigg],
\end{aligned}
\end{equation}
where the abbreviations \(x_{\tq,\bq}\) and \(y_{\tq,\bq}\) are
\begin{subequations}\label{eq:def-xy-gen}
\begin{align}
x_\tq^2 &= \frac{\left|A_\tq h_\uq^0 - \mu^* Y_\tq
    h_\dq^{0*}\right|^2}{{\widetilde{M}}^4_\tq} + \frac{\left({\tilde m}^2_Q -
    {\tilde m}^2_t\right)^2}{4 {\widetilde{M}}^4_\tq}, \quad
y_\tq = \frac{\left|Y_\tq h_\uq^0\right|^2}{{\widetilde{M}}_\tq^2}, \\
x_\bq^2 &= \frac{\left|A_\bq h_\dq^0 - \mu^* Y_\bq
    h_\uq^{0*}\right|^2}{{\widetilde{M}}_\bq^4} + \frac{\left({\tilde m}^2_Q -
    {\tilde m}^2_b\right)^2}{4 {\widetilde{M}}^4_\bq}, \quad
y_\bq = \frac{\left|Y_\bq h_\dq^0\right|^2}{{\widetilde{M}}_\bq^2}.
\end{align}
\end{subequations}
The soft SUSY breaking masses enter as \(\tilde m_Q^2\), \(\tilde
m_t^2\) and \(\tilde m_b^2\) and we defined \(\widetilde M_{\tq,\bq}^2 =
(\tilde m_Q^2 + \tilde m_{t,b}^2) / 2\). The trilinear soft breaking
couplings in the up and down sector are given by \(A_\tq\) and
\(A_\bq\), respectively. Yukawa couplings are denoted as \(Y_{\tq,\bq}\)
and \(\mu\) is the \(\mu\) parameter of the superpotential in the MSSM.
The mass parameters of the tree-level Higgs potential are
\(m_{11}^{2\,^\text{tree}} = m_{H_\dq}^2 + |\mu|^2\),
\(m_{22}^{2\,^\text{tree}} = m_{H_\uq}^2 + |\mu|^2\) and
\(m_{12}^{2\,^\text{tree}} = B_\mu\) with the soft breaking masses
\(m_{H_\uq}^2\) and \(m_{H_\dq}^2\) for the \(H_\uq\) and \(H_\dq\)
doublet, respectively; \(B_\mu\) is the soft breaking bilinear term
\(\sim H_\uq \cdot H_\dq\). We consider only third generation
superfields which couple with large Yukawa couplings to the Higgs
doublets:
\begin{equation}\label{eq:superpot}
\mathcal{W} = \mu\; H_\dq \cdot H_\uq + Y_\tq\; H_\uq \cdot Q_\Ll \bar T_\Rr
- Y_\bq\; H_\dq \cdot Q_\Ll \bar B_\Rr.
\end{equation}
The left-handed doublet field is \(Q_\Ll = ( T_\Ll, B_\Ll )\) and the
two Higgs doublets \(H_\uq = ( h_\uq^+, h_\uq^0 )\) and \(H_\dq = (
h_\dq^0, -h_\dq^-)\); \(\SU(2)_\Ll\)-invariant multiplication is denoted
by the dot-product. The \(\SU(2)_\Ll\) singlets are put into the
left-chiral supermultiplets \(\bar T_\Rr = \{ \tilde t_\Rr^*,
t^c_\Rr\}\) and \(\bar B_\Rr = \{ \tilde b_\Rr^*, b_\Rr^c\}\) with the
charge conjugated Weyl spinors \(t^c_\Rr\) and \(b_\Rr^c\).

The effective potential of Eq.~\eqref{eq:VeffMSSM} obviously develops an
imaginary part beyond the branch point of the logarithms
\(\ln(1\pm x+y)\). We want to give a physical meaning of this branch point
without reference to an imaginary part of the effective potential, since
\(\frac{1}{2}\ln\big((1\pm x+y)^2\big)\) does not reveal any imaginary
part---nevertheless, this logarithm gets singular where \(\mp x-y=1\) though
the potential itself stays finite. This point determines (for fixed
parameters) a critical Higgs field value for which one mass eigenvalue
gets tachyonic. The effective potential is a function of the (classical)
field values which correspond to vacuum expectation values at the
minimum. In the direction of the negative mass square, the potential
drops down and therefore develops a CCB vacuum.

Moreover, for certain parameters, the potential of
Eq.~\eqref{eq:VeffMSSM} develops a second minimum in the direction of a
standard Higgs \vev{} which always lies beyond the branch point of one
of the logarithms~\cite{Bobrowski:2014dla}. Expanding around this second
minimum, one finds exactly one negative sbottom mass square (in the
region of large \(\mu\) and \(\tan\beta\)) which hints towards a global
minimum including a sbottom \vev{}. The second minimum as depicted
in~\cite{Bobrowski:2014dla} is an artifact of holding \(\tilde b_{\Ll,
  \Rr} = 0\): the global minimum lies at a point with both \( \langle
\tilde b_{\Ll, \Rr} \rangle \neq 0\) and \(\langle h_\uq^0 \rangle \neq
v_\uq\).

We take the existence of the critical field value serious and first
figure out its meaning for the development of such a CCB minimum. For
simplification we now restrict ourselves in the following to \(\langle
\tilde t_L \rangle = \langle \tilde t_R \rangle = 0\) and also do not
consider stau \vev{}s. Let us consider for the moment a fixed value of
the down-type Higgs field, \(h_\dq^0 = v_\dq\) and set \(A_\bq =
0\). The critical field value is then obtained by solving \(x_\bq -
y_\bq = 1\) with \(x_\bq\) and \(y_\bq\) given in
Eq.~\eqref{eq:def-xy-gen}:
\begin{equation}\label{eq:hcrit}
  h^0_\uq \big|_\text{crit} = \pm \frac{ Y_\bq^2 v_\dq^2 +
      M_\text{SUSY}^2}{\mu Y_\bq},
\end{equation}
with \(\tilde m_Q^2 = \tilde m_b^2 = M_\text{SUSY}^2\) and \(\mu\),
\(Y_\bq\) as well as the Higgs field assumed to be real. The bottom
Yukawa coupling suffers from SUSY threshold corrections and reads
\(Y_\bq = m_\bq / [v_\dq (1+\Delta_b)]\) with \(\Delta_b\) including the
Higgsino corrections \(\sim \mu A_\tq \tan\beta\)~\cite{Hall:1993gn,
  Carena:1994bv, Pierce:1996zz, Carena:1999py}, which can be dominant
over the gluino-induced threshold correction for large \(\mu\tan\beta\)
and large gluino mass. Both gluino and higgsino contributions sum up
together, \(\Delta_b = \Delta_b^\text{gluino} +
\Delta_b^\text{higgsino}\), where the interesting one-loop contribution
is given by~\cite{Hall:1993gn, Carena:1994bv, Pierce:1996zz,
  Carena:1999py}
\begin{subequations}\label{eq:Deltab}
\begin{align}
  \Delta_b^\text{gluino} &= \frac{2\alpha_s}{3\pi} \mu M_{\tilde G}
  \tan\beta I(\tilde m_{\tilde b_1}, \tilde m_{\tilde b_2}, M_{\tilde
    G}
  ), \label{eq:Deltabglu} \\
  \Delta_b^\text{higgsino} &= \frac{Y_\tq^2}{16\pi^2} \mu A_\tq
  \tan\beta I(\tilde m_{\tilde t_1}, \tilde m_{\tilde t_2},
  \mu), \label{eq:Deltabhig}
\end{align}
\end{subequations}
with
\[
I(m_1, m_2, m_3) = \frac{
  m_1^2 m_2^2 \ln \frac{m_2^2}{m_1^2} +
  m_2^2 m_3^2 \ln \frac{m_3^2}{m_2^2} +
  m_1^2 m_3^2 \ln \frac{m_1^2}{m_3^2}
}{(m_1^2 - m_2^2)(m_1^2 - m_3^2)(m_2^2 - m_3^2)}.
\]
There are also higher order calculations of \(\Delta_b\) available that
are important for precision analyses~\cite{Hofer:2009xb,
  Crivellin:2011jt, Crivellin:2012zz}.

The gluino loop contribution~\eqref{eq:Deltabglu} decouples with the
gluino mass if the other SUSY parameters are fixed, but the higgsino
one~\eqref{eq:Deltabhig} cannot be neglected for the desired values of
\(\mu\) around the SUSY scale. For the numerical analysis in the course
of this letter, we set \(M_{\tilde G} = M_\text{SUSY}\) which reduces
\(Y_\bq\) for positive \(\mu\). Moreover, we only include ``active''
third generation squarks as superpartners and implicitly take any other
superpartner heavy (all gauginos besides the gluino which does not give
a contribution to the effective Higgs potential at one-loop).

There are handy exclusion limits, well-known for a long time, to simply
check whether an unwanted, charge and color breaking (CCB) minimum
appears for a given set of parameters in the MSSM. The constraints are
on soft breaking trilinear couplings against soft breaking mass
parameters as
\begin{equation}\label{eq:tradCCB}
A_\tq^2 < 3 (m_{22}^2 + {\tilde m}_Q^2 + {\tilde m}_t^2 ),
\end{equation}
see \eg \cite{Frere:1983ag, Claudson:1983et, Derendinger:1983bz,
  Kounnas:1983td, AlvarezGaume:1983gj, Ibanez:1983di, Drees:1985ie,
  Casas:1995pd}.

Mostly studied, however, are such couplings of up-type squarks to the
up-type Higgs or of down-type sleptons to the down-type Higgs (where
similar expression for down-type squarks can be obtained by relabeling
the parameters). Couplings to the ``wrong'' Higgs doublet are mainly
excluded in the analyses. The destabilizing contribution is always
related to the trilinear part of the scalar potential, \eg \(\sim \mu
Y^*_\bq h_\uq^0 \tilde{b}^*_\Rr \tilde b_\Ll\). It has been
shown~\cite{Bobrowski:2014dla} that the direction of the up-type Higgs
field gets apparently destabilized from a (s)bottom loop
effect. In~\cite{Bobrowski:2014dla} only the field direction of the
neutral Higgs, \(h_\uq^0\), was considered---we now want to give a more
complete view of the destabilizing effect leading to an analytic
approximate exclusion on the combination \(\mu Y_\bq\) in case the
colored sbottom direction is included. Another exclusion can be obtained
using a different direction in field space, where also the down-type
Higgs scalar is needed.

In this letter, we describe in the following section how to derive the
analytic expression for the new CCB constraint from sbottom \vev{}s and
compare it to the numerical analysis of the global minima in the quantum
(\eg loop corrected) theory. Finally, we conclude.

\section{Finding CCB minima}
So far, we only discussed features of the scalar (one-loop) Higgs
potential from Eq.~\eqref{eq:VeffMSSM} as described
in~\cite{Bobrowski:2014dla}. In order to find the new (true) CCB vacuum,
which hides behind the critical Higgs field value, we add to the
potential of Eq.~\eqref{eq:VeffMSSM} (evaluated at \(Q^2 =
M_\text{SUSY}^2\)) the \emph{tree-level part} of the sbottom potential,
\begin{equation}\label{eq:Vbtree}
\begin{aligned}
V^\upsh{tree}_{\tilde b} &= \tilde b_\Ll^* ({\tilde m}_Q^2 + |Y_\bq
h^0_\dq|^2) \tilde b_\Ll + \tilde b_\Rr^* ({\tilde m}_b^2 + |Y_\bq
h^0_\dq|^2) \tilde b_\Rr \\
&\quad - \left[ \tilde b_\Ll^* (\mu^* Y_\bq h_\uq^{0\dag} - A_\bq
h^0_\dq) \tilde b_\Rr + \hc \right] + |Y_\bq|^2 |\tilde b_\Ll|^2 |\tilde
b_\Rr|^2 + \;D\text{-terms}.
\end{aligned}
\end{equation}
As was already pointed out before~\cite{Gunion:1987qv, Rattazzi:1996fb},
the destabilizing term is always the trilinear one, \(\mu Y_\bq^*
h_\uq^0 \tilde b^*_\Rr \tilde b_\Ll\), so we expect a new stability
condition for the combination \(\mu Y_\bq\) taking \(A_\bq =
0\). Actually, we cannot ignore \(D\)-terms in the tree-level potential
accordingly to the neglect of all \(g_{1,2}^2\) terms in the derivation
of the one-loop Higgs potential, since also the Higgs self-couplings are
\(\sim g_{1,2}^2\). However, we can simplify (as usually done) the
discussion considering so-called ``\(D\)-flat'' directions. Those
directions are most probably that kind of rays in field space in which
unwanted minima develop. Non-\(D\)-flat directions are protected by the
quartic terms that will always take over. The full \(D\)-term potential
for the Higgs and sbottom scalar potential is given by
\begin{equation}\label{eq:VDfull}
\begin{aligned}
  V_D =&\; \frac{g_1^2}{8} \big( |h_\uq^0|^2 - |h_\dq^0|^2 + \frac{1}{3}
  |\tilde b_L|^2 + \frac{2}{3} |\tilde b_R|^2 \big)^2 \\
  +&\; \frac{g_2^2}{8} \big( |h_\uq^0|^2 - |h_\dq^0|^2 + |\tilde b_L|^2
  \big)^2 + \frac{g_3^2}{6} \big(|\tilde b_L|^2 - |\tilde b_R|^2\big)^2.
\end{aligned}
\end{equation}
We still ignore stop and stau fields and remark that the pure Higgs
terms are already included in Eq.~\eqref{eq:VeffMSSM}. Nevertheless, we
make use of Eq.~\eqref{eq:VDfull} to set the interesting directions:
with \(\tilde b_L = \tilde b_R \equiv \tilde b\), we have the
\(\SU(3)_c\) \(D\)-flat direction. Considering the three-field scenario,
we can reduce the degrees of freedom forcing all \(D\)-terms to vanish
by the choice \(|h_\dq^0|^2 = |h_\uq^0|^2 + |\tilde b|^2\). Still rather
large quartic terms survive in the potential, namely the \(|Y_\bq|^2\)
terms from the \(F\)-term part in \(V_{\tilde b}^\text{tree}\). For that
observation, we also look into a non-\(D\)-flat direction keeping
\(\frac{g_1^2 + g_2^2}{8} \big( |h_\uq^0|^2 + |\tilde b|^2 \big)^2\),
where the down-type Higgs is fixed at \(h_\dq^0 = v_\dq\) which is a
constant and small number especially for large
\(\tan\beta\),\footnote{With \(v_\dq = v \cos\beta\) we denote the
  standard electroweak \vev{} of the down-type Higgs.} and therefore
neglected with respect to potentially large field values of \(\tilde b\)
and \(h_\uq^0\). Note that contrary to most previous considerations
\cite{Gunion:1987qv, Casas:1995pd, Kusenko:1996jn, Chowdhury:2013dka} we
are explicitly interested in \(\tilde b \neq 0\) though \(A_\bq = 0\)
and have \(\tilde t_{\Ll, \Rr} = 0\). In both ways we are considering a
combined non-standard vacuum in the mixed sbottom and \emph{up}-type Higgs
direction instead of the pure down-down case.

Let us figure out the analytic bound analogously to the famous
\(A\)-parameter bounds like Uneq.~\eqref{eq:tradCCB}, under which
circumstances a CCB \emph{true} vacuum appears. For that purpose, we
shall choose the most probable field configuration that makes all the
\(D\)-terms vanish. In the \(\SU(3)_c \times \SU(2)_\Ll \times
\U(1)_Y\) \(D\)-flat direction, we assign \(\tilde b_\Ll = \tilde
b_\Rr = \tilde b\) and \({h^0_\dq}^2 = {h_\uq^0}^2 + \tilde b^2\). We
consider only real fields and parameters now and in the following for
simplicity. A different but not uninteresting bound will be derived in
a direction where we keep the \(h_\dq\) field strength at a fixed and
small value, \(h_\dq^0 = v_\dq \approx 0\). That way, we cannot reduce
the quartic terms but still find a (new) analytic exclusion in the
\(h_\uq^0 = \tilde b\) direction.

\begin{figure}[t]
\begin{minipage}{.5\textwidth}
\includegraphics[width=\textwidth]{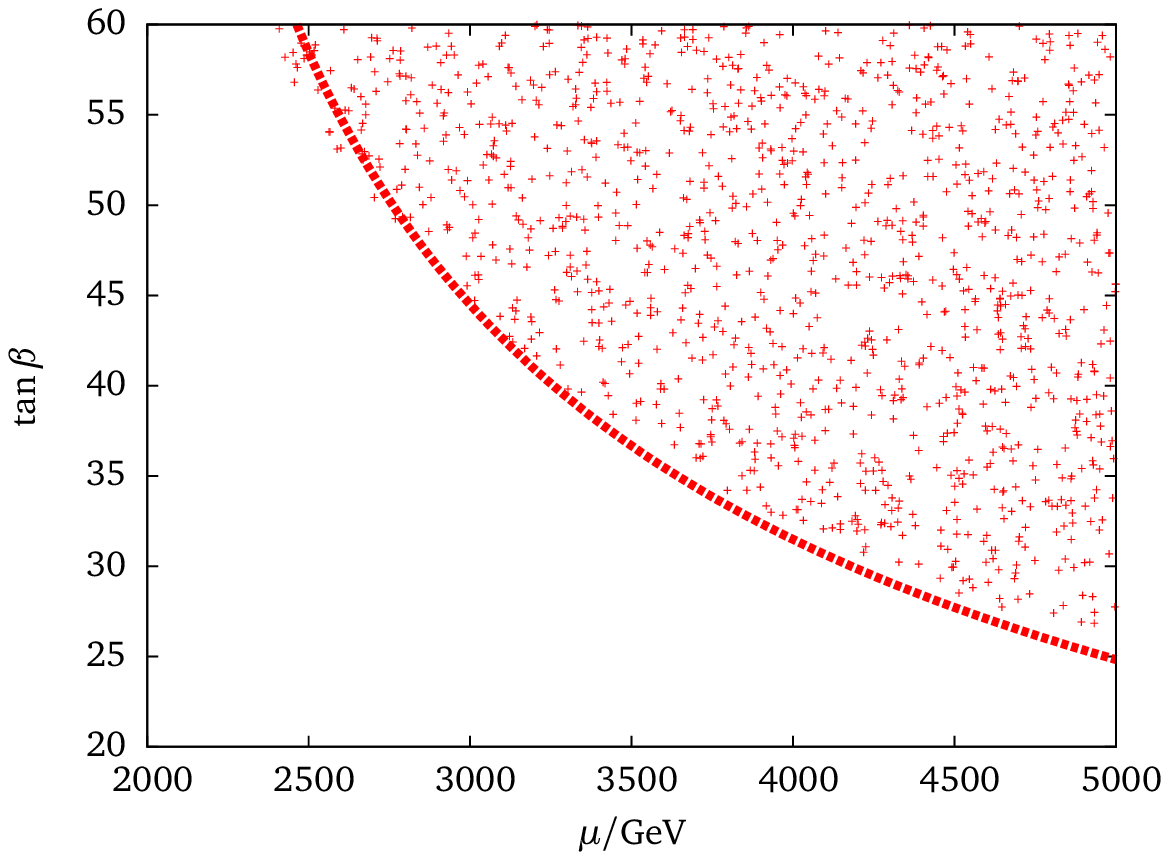}
\end{minipage}%
\begin{minipage}{.5\textwidth}
\includegraphics[width=\textwidth]{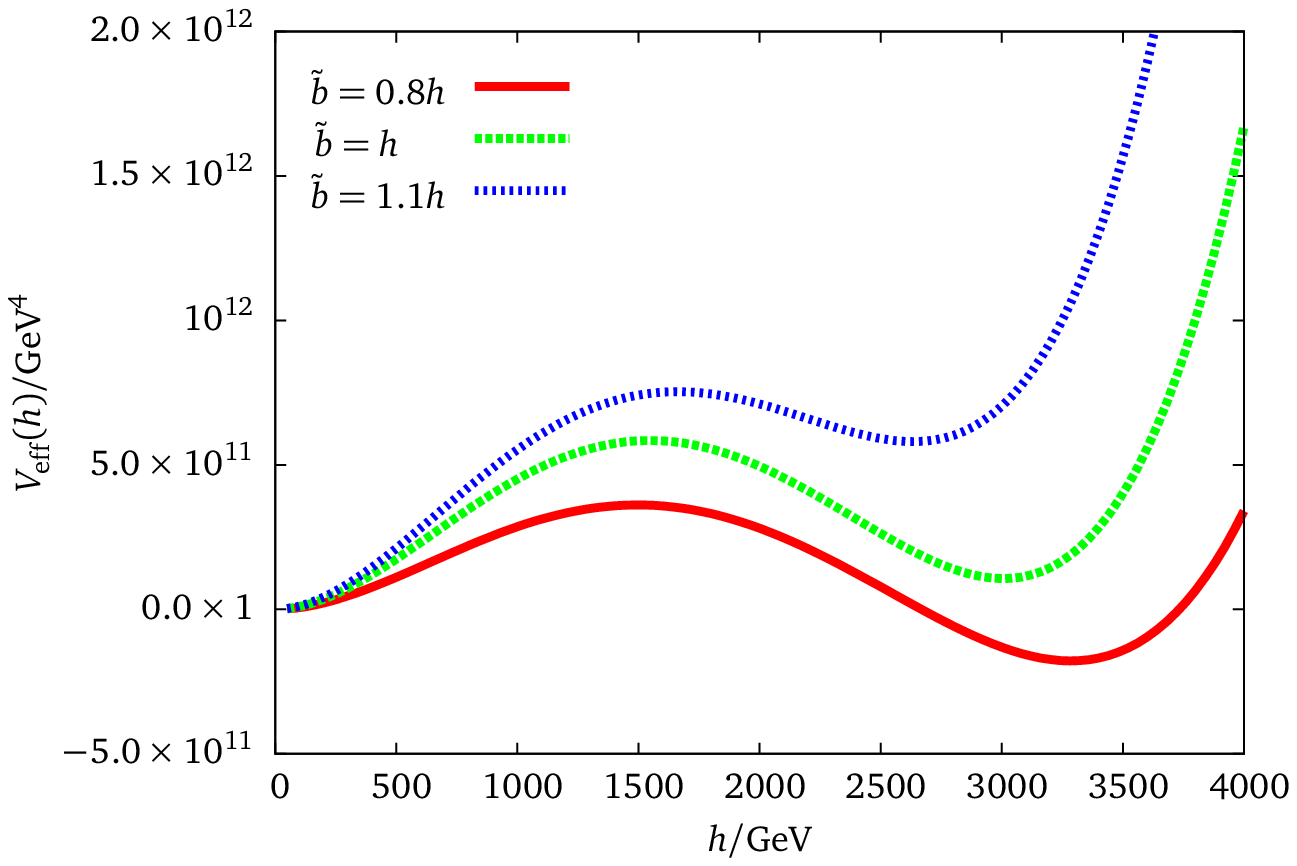}
\end{minipage}
\caption{New exclusion limits including the formation of charge and
  color breaking minima with both \(\langle \tilde b \rangle \neq 0\)
  and \(\langle h_\uq^0 \rangle \neq 0\). The left plot shows
  exclusions in the \(\mu\)-\(\tan\beta\) plane with \(A_\tq = -
  1800\,\GeV\) which has been chosen conveniently to give the proper
  light Higgs mass within a few \(\GeV\) and a common soft breaking
  mass \(M_\text{SUSY} = 1\,\TeV\). The small points are excluded by
  the numerical comparison of the two minima (exclusion if CCB minimum
  deeper than trivial one). On the right-hand side we depict the
  crucial dependence of the non-standard minimum on the (mis)alignment
  of sbottom field and Higgs field value (\(\tilde b = \alpha h\) with
  \(h = h_\uq^0\), \(h_\dq^0=0\) and \(\alpha \in \{ 0.8, 1, 1.1\}\)
  for a given excluded point.}
\label{fig:CCBexcl-nonflat}
\hrule
\end{figure}

\paragraph{\(h_\uq^0 = \tilde b\)}
An exact analytic derivation of the exclusion limits from the
stability of the electroweak vacuum against formation of charge and
color breaking minima is very easy to obtain in the one-field
scenario. We follow the standard procedure which was
pictorially reviewed in Ref.~\cite{Camargo-Molina:2013sta}. We collect
the interesting parts of the tree-level potentials of
Eqs.~\eqref{eq:VeffMSSM} and \eqref{eq:Vbtree},
\begin{equation}\label{eq:Vhbtree}
  V_{\tilde b, h}^\text{tree} = (M^2 - 2 \mu Y_\bq h) \tilde b^2 +
  m^2 h^2 + \lambda_b \tilde b^4 + \lambda_h h^4 + \lambda_{hb} h^2 \tilde
  b^2,
\end{equation}
with \(M^2 = {\tilde m}_Q^2 + {\tilde m}_b^2\), \(m^2 = m_{H_\uq}^2 +
\mu^2\) and the self-couplings \(\lambda_b = Y_\bq^2 +
\frac{g_1^2+g_2^2}{8}\), \(\lambda_h = \frac{g_1^2+g_2^2}{8}\) and
\(\lambda_{hb} = \frac{g_1^2 + g_2^2}{4}\). This simplifies via \(\tilde
b = h\) further to
\begin{equation}\label{eq:Vhhtree}
V_{h,h}^\text{tree} = \bar{m}^2 h^2 - A h^3 + \lambda h^4,
\end{equation}
with \(\bar{m}^2 = M^2 + m^2\), \(\lambda = \lambda_h + \lambda_b +
\lambda_{hb}\) and \(A = 2 \mu Y_\bq\). We then find with the \vev{},
\[v = \langle h \rangle = \frac{3 A + \sqrt{9 A^2 - 32 \bar{m}^2
    \lambda}}{8 \lambda}, \] and the requirement\footnote{The potential
  of Eq.~\eqref{eq:Vhhtree} reveals a strong first order phase
  transition, where the trivial minimum appears to be
  \(V(h=0)=0\). Stable configurations need the potential value to be
  larger than that one.}  that for stable configurations \(V_\text{min}
= V_{h,h}^\text{tree} (v) > 0\), which is \(\bar{m}^2 > \frac{A^2}{4
  \lambda}\), the new condition as (\(m_{H_\uq}^2\) is negative!)
\begin{equation}\label{eq:analyticexcl}
m_{H_\uq}^2 + \mu^2 + {\tilde m}_Q^2 + {\tilde m}_b^2 >
\frac{( \mu Y_\bq)^2 }{Y_\bq^2 + (g_1^2 + g_2^2)/2}.
\end{equation}
Note that \(Y_\bq\) has a non-trivial dependence on \(\mu\),
\(\tan\beta\) and also \(A_\tq\) via \(\Delta_b\), see
Eqs.~\eqref{eq:Deltab} and~\cite{Hall:1993gn, Carena:1994bv,
  Pierce:1996zz, Carena:1999py}. The \((g_1^2 + g_2^2)/2\)
contribution is the left-over from the non-\(D\)-flatness which can be
numerically of the same size as a threshold-resummed \(Y_\bq\),
weakening the exclusion. This bound, however, does not fit exactly to
the numerical exclusion as can be seen from
Fig.~\ref{fig:CCBexcl-nonflat} but provides an excellent approximation
though actually \(\langle h\rangle \neq \langle \tilde b\rangle\). The
numerical exclusion limit shown in Fig.~\ref{fig:CCBexcl-nonflat}
agrees well with independent previous analyses on a similar
situation~\cite{Altmannshofer:2012ks} and are a bit stricter than the
final results of~\cite{Bobrowski:2014dla}, whereas a similar necessary
condition was found for a slightly different direction in field
space~\cite{Altmannshofer:2014qha}.

\begin{figure}[t]
\begin{minipage}{.49\textwidth}
\includegraphics[width=\textwidth]{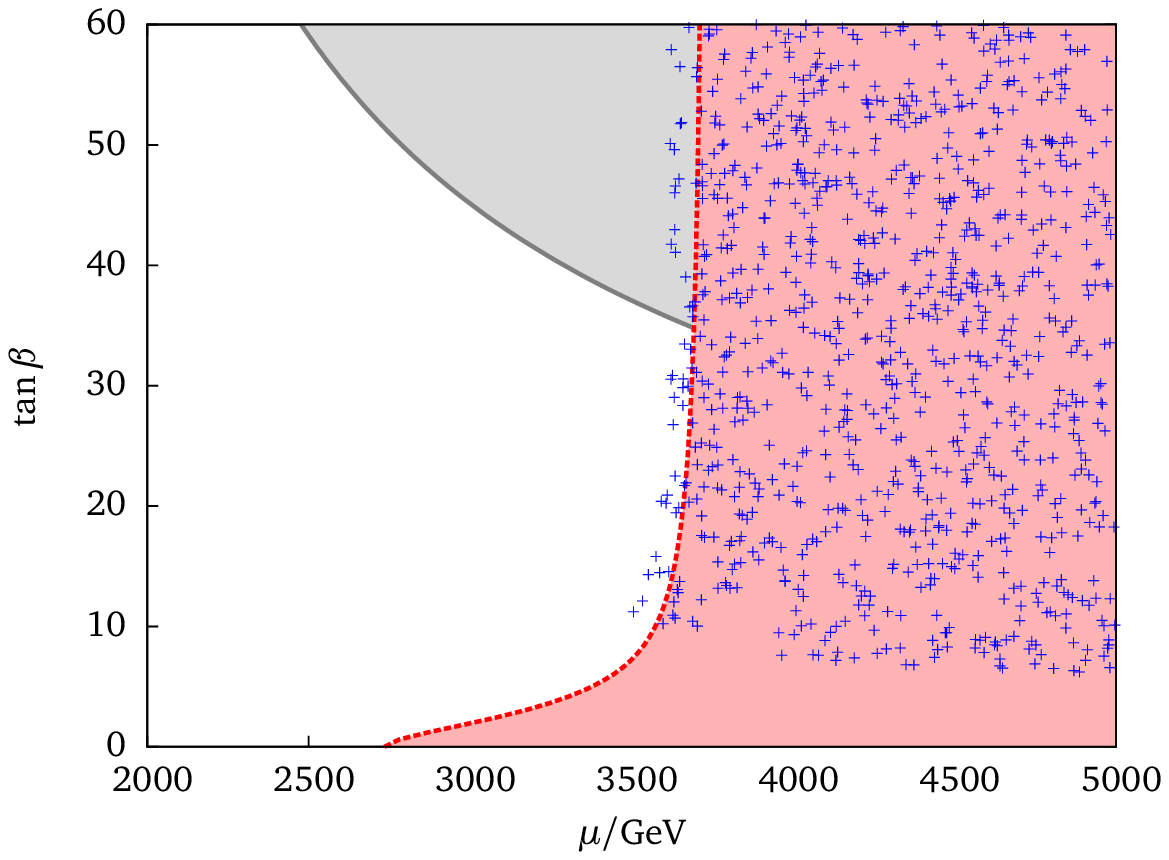}
\end{minipage}%
\hfill
\begin{minipage}{.49\textwidth}
\includegraphics[width=\textwidth]{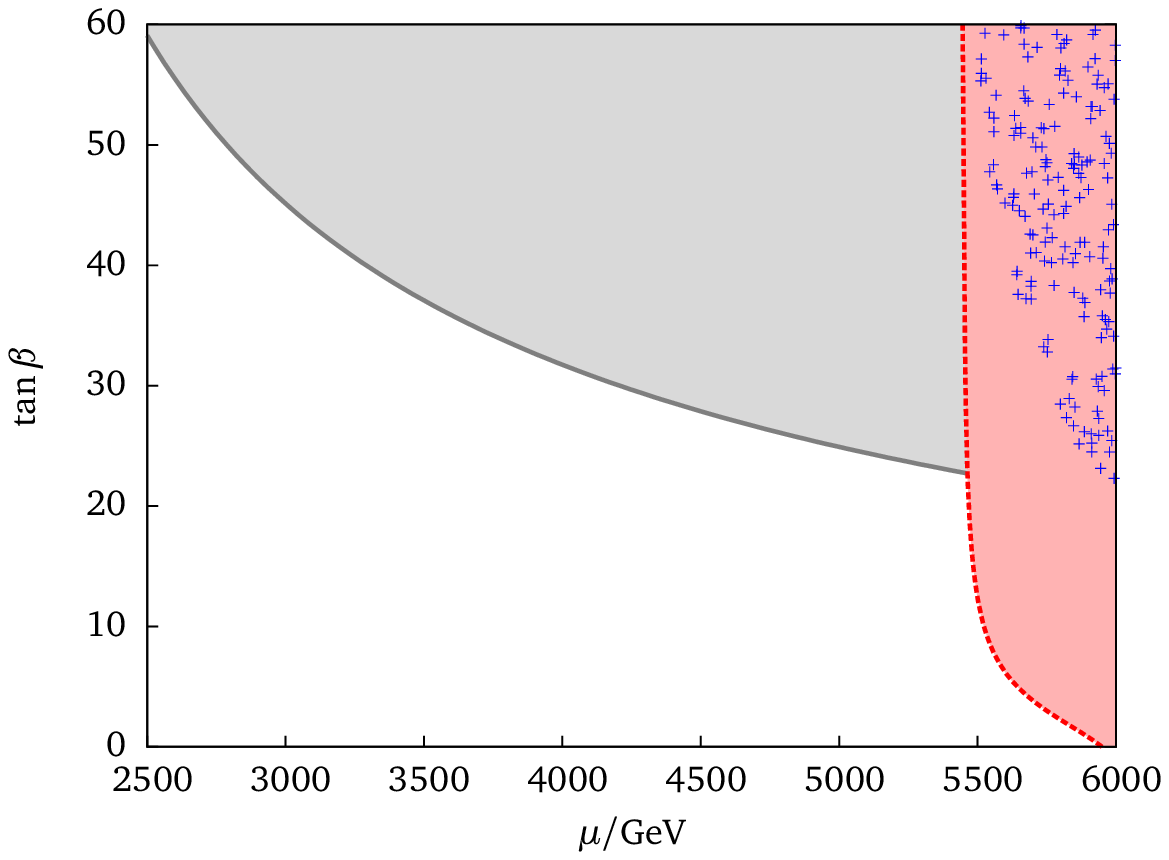}
\end{minipage}
\caption{Exclusion in the \(\mu\)-\(\tan\beta\) plane similar to the
  one shown in Fig.~\ref{fig:CCBexcl-nonflat} (which is indicated by
  the grayish area) for the \(D\)-flat direction \(|h_\dq^0|^2 =
  |h_\uq^0|^2 + |\tilde b|^2\). Blue dots have been excluded via
  numerical comparison of the two minima (if so) using the one-loop
  Higgs potential and an improved sbottom potential at the tree-level;
  the red line shows the exclusion of Uneq.~\eqref{eq:flatbound} where
  the misalignment parameter \(\alpha\) has been ``fitted'' for
  optical agreement of the blue dots and the reddish area to be
  \(0.75\); the actual \(\alpha\) are different for each blue
  point. On the left-hand side, we have the \(-\)-sign and on the
  right-hand side the \(+\)-sign of Uneq.~\eqref{eq:flatbound}.}
\label{fig:CCBexcl-flat}
\hrule
\end{figure}

\begin{figure}[t]
\begin{minipage}{.49\textwidth}
\includegraphics[width=\textwidth]{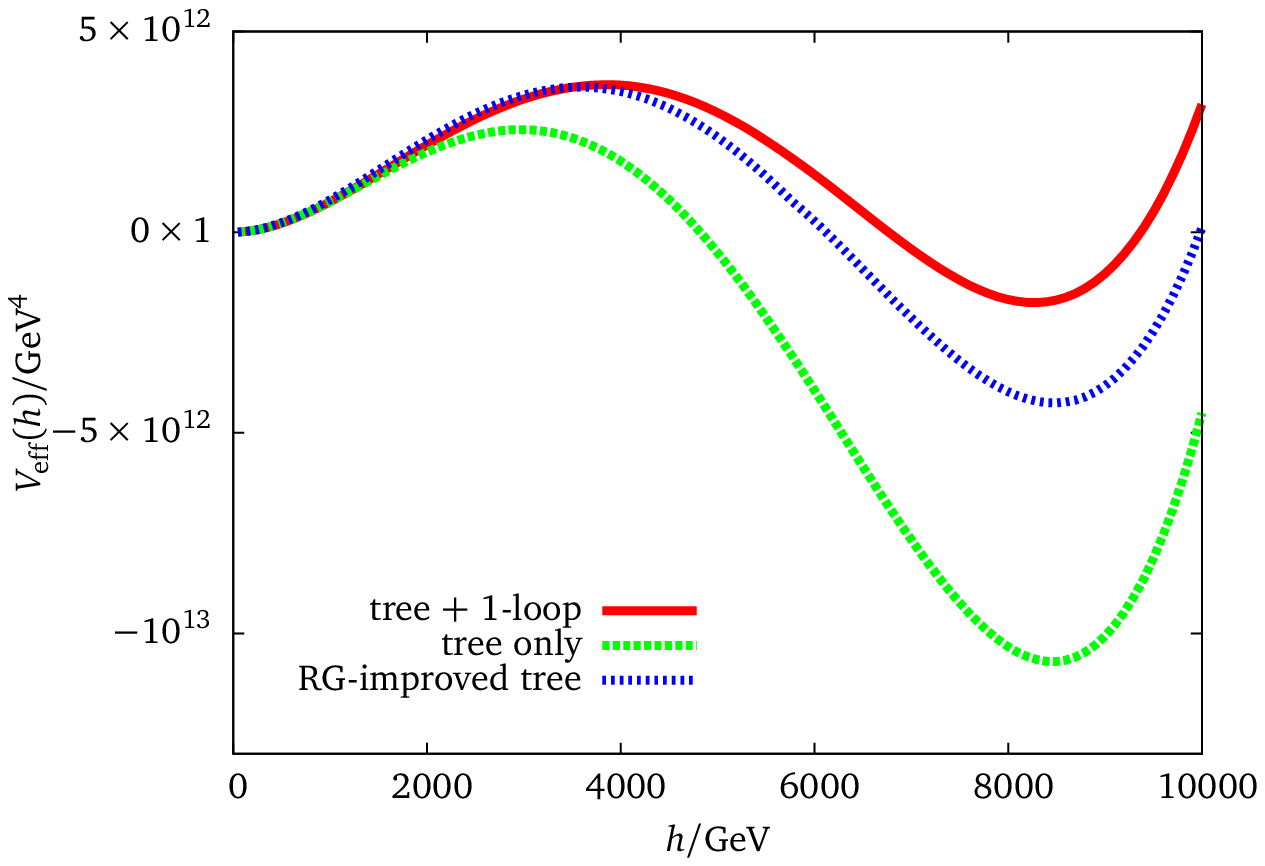}
\end{minipage}%
\hfill
\begin{minipage}{.49\textwidth}
\includegraphics[width=\textwidth]{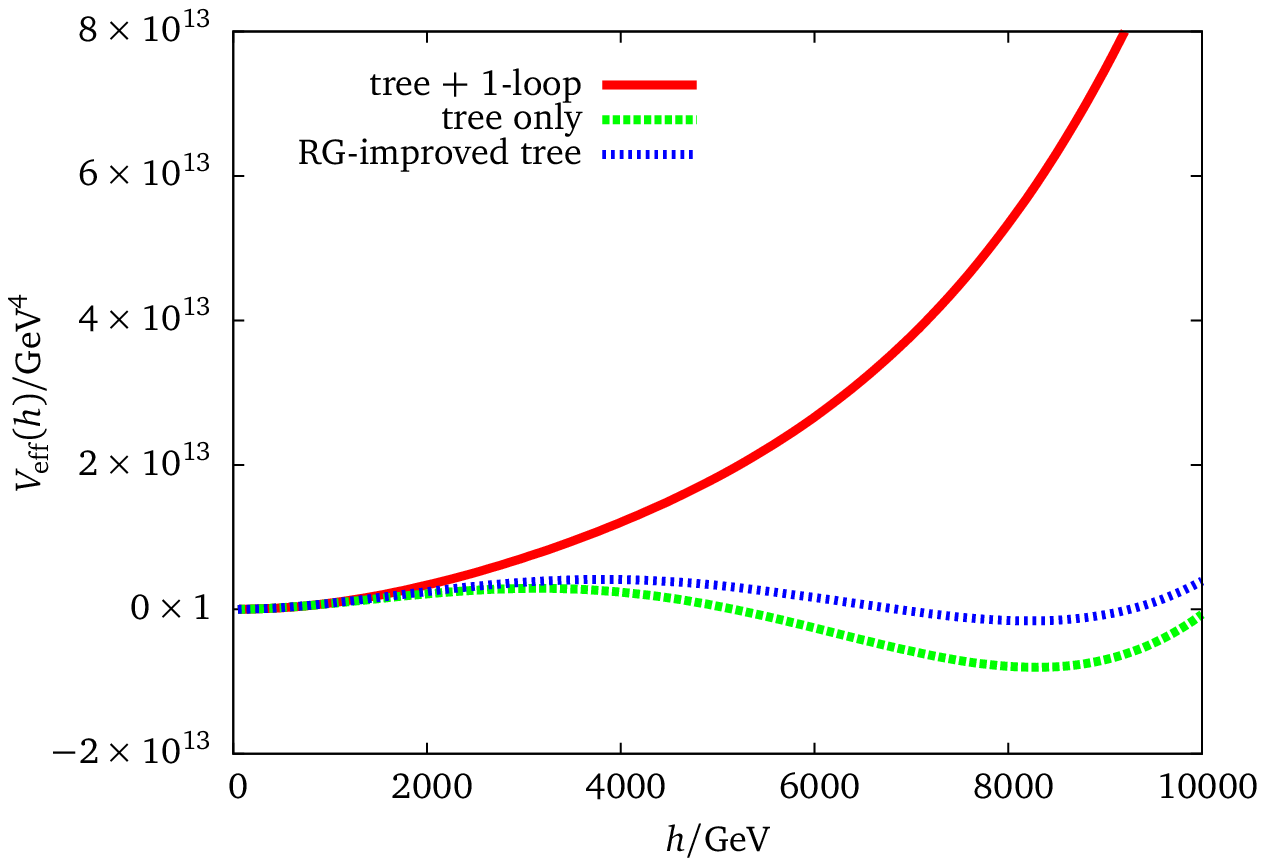}
\end{minipage}
\caption{For a given parameter point (\(\mu=4\,\TeV\),
  \(\tan\beta=40\), \(A_\tq = 1.8\,\TeV\)), we show exemplarily the
  behavior of the potential in the given direction in field space (and
  \(\tilde b = 0.75 h_\uq^0\)). On the left-hand side, the positive
  sign for \(h_\dq^0\) was chosen, where the plot on the right has
  \(h_\dq^0 = - \sqrt{|h_\uq^0|^2 + |\tilde b|^2}\) with real fields
  and parameters in both cases. The ``tree + 1-loop'' line means
  inclusion of the one-loop Higgs potential as of
  Eq.~\eqref{eq:VeffMSSM} plus the tree-level bottom squark potential
  (without \(D\)-terms each since they vanish by definition of the
  direction) evaluated with running parameters (soft-breaking squark
  masses and Yukawa couplings). For comparison, we show the ``tree
  only'' where the masses and couplings of the potential have been
  evaluated at the SUSY scale \(M_\text{SUSY} = 1\,\TeV\) and the
  ``RG-improved tree'' potential where all soft masses and couplings
  are treated as running ones.}
\label{fig:CCBpot-flat}
\hrule
\end{figure}

\paragraph{\(|h^0_\dq|^2 = |h_\uq^0|^2 + |\tilde b|^2\)}
With the knowledge from above, it is straightforward to give a similar
exclusion in the \(D\)-flat direction \(|h^0_\dq|^2 = |h_\uq^0|^2 +
|\tilde b|^2\). The remaining two-field scalar potential (real
fields and parameters, \(A_\bq = 0\)) can be further reduced aligning
\(\tilde b = \alpha h_\uq^0 = \alpha h\) with a (real) scaling
parameter \(\alpha\):
\begin{equation}
\begin{aligned}
V_{D\text{-flat}} &= \left( m_{11}^2 (1 + \alpha^2) + m_{22}^2 \pm 2
  m_{12}^2 \sqrt{1+\alpha^2} + \alpha^2 ({\tilde m_Q}^2 + {\tilde
    m_b}^2) \right) h^2 \\ &\;
  - 2 \mu Y_\bq \alpha^2 h^3 + Y_\bq^2 \big(2 \alpha^2 ( 1 + \alpha^2 ) + \alpha^4 \big) h^4,
\end{aligned}
\end{equation}
that can be easily mapped on the expression of Eq.~\eqref{eq:Vhhtree}
resulting in the requirement that for stable
configurations\footnote{The sign ambiguity origins from the fact, that
  we only need to constrain \(|h_\dq^0|^2\) where the overall phase or
  sign is not constrained.}
\begin{equation}\label{eq:flatbound}
 m_{11}^2 (1 + \alpha^2) + m_{22}^2 \pm 2
  m_{12}^2 \sqrt{1+\alpha^2} + \alpha^2 ({\tilde m_Q}^2 + {\tilde
    m_b}^2) > \frac{\mu^2 \alpha^2}{2 + 3 \alpha^2}.
\end{equation}
This exclusion translated into the \(\mu\)-\(\tan\beta\) plane is
shown in Fig.~\ref{fig:CCBexcl-flat} where we also display points that
are excluded via the numerical minimization of the combined tree and
one-loop effective potential. To enhance the significance of this
bound (which is basically \(\tan\beta\)-independent), we have employed
running squark parameters in the tree-level sbottom potential
evaluated at the scale of the new minimum. Therefore, also
corresponding parameters in the analytic exclusion (soft SUSY breaking
masses and \(\mu\)) have been taken at the same scale. Unfortunately,
for the purpose of displaying the exclusion line, it is not clear at
which scale those parameters have to be evaluated. As the second
minimum generically appears around one order of magnitude above the
SUSY scale, we have set a fixed renormalization scale of \(10\,
M_\text{SUSY}\) and therefore blue dots and the reddish area on the
left-hand side Fig.~\ref{fig:CCBexcl-flat} do not perfectly
fit. Moreover, the excluded area by Uneq.~\eqref{eq:flatbound} is not
completely filled with excluded blue points as there the sbottom-tree
plus Higgs-one-loop potential shows a different behavior than the
classical potential as also depicted in Fig.~\ref{fig:CCBpot-flat}.

Unequations like~\eqref{eq:flatbound} or~\eqref{eq:analyticexcl}
follow from the tree-level potential and can be determined easily once
a specific field line is selected. Going beyond tree-level changes the
situation severely as can be seen from Fig.~\ref{fig:CCBpot-flat}. A
configuration which is obviously unstable ({\color{red} right}-hand
side) at the tree-level not even develops a second minimum considering
the one-loop Higgs potential (the complete one-loop potential
including sbottom directions was not employed for that purpose though
should be available numerically). However, this effect is different in
the ``positive'' \(h_\dq^0\) direction where unstable configurations
are driven towards more stable ones as can be seen from the left-hand
side of Fig.~\ref{fig:CCBpot-flat}. Usage of the \emph{renormalization
  group improved} (tree-level) effective potential, where the
couplings (Yukawa couplings and masses, actually no gauge couplings
are they are absent in the genuine \(D\)-flat direction) are evaluated
at a proper scale,\footnote{The choice of a proper renormalization
  scale is a bit vague and the decision whether to trust that choice
  in order to discard certain configurations is tenuous. For our
  purpose, we stick to the suggestion of Ref.~\cite{Gamberini:1989jw}
  and choose a scale \(\hat Q = \operatorname{max} \left(
    \mathcal{M}_f^2(h) \right)\) as the largest field-dependent mass
  eigenvalue of the loop-contributing fields (in our case top and/or
  bottom (s)quark).} hint towards less restrictive exclusions. Where
the tree-only potentials show non-trivial charge and color breaking
minima, the loop-corrected potentials seem to stabilize the standard
vacuum against formation of false vacua.

\paragraph{Estimate of lifetime}
Are the developing charge and color breaking minima really a case for
anxiety? As long as the lifetime of the ``standard'' electroweak
vacuum is (much) longer than the present age of the universe, we
basically do not have not worry and can take the issue of vacuum
metastability for future generations. We estimate the lifetime of the
desired vacuum for the scenarios provided in
Figs.~\ref{fig:CCBexcl-nonflat} and \ref{fig:CCBpot-flat} using the
triangle method of~\cite{Duncan:1992ai} and the instable potentials
shown in the figures. However, similar to the scenario discussed
in~\cite{Bobrowski:2014dla}, where the decay time was found to be
ridiculously small (details on the estimate have been given
in~\cite{Hollik:2015lwa}), we find our unstable solutions to be
extremely short-lived concerning Fig.~\ref{fig:CCBexcl-nonflat}. This
is not true for the genuine \(D\)-flat scenario shown in
Fig.~\ref{fig:CCBpot-flat}; here the lifetime is many orders the
lifetime of the universe.

\section{Conclusions}
We have provided new (analytic) exclusion bounds in the MSSM from the
formation of CCB minima. Contrary to previous considerations, we did
not constrain the soft-breaking \(A\)-parameter by working in the
direction of up or down fields only but connected the bottom squark
direction with the up-type Higgs field. This procedure gives a
constraint on \(\mu Y_\bq\), where the bottom Yukawa coupling has an
implicit dependence on the model parameters via \(Y_\bq = m_\bq /
[v_\dq(1 + \Delta_b)]\). Under certain simplifications we have derived
an analytic bound which is mostly in good agreement with the direct
numerical exclusion from the minimization of the full (\ie tree-level
sbottom plus one-loop Higgs) effective potential considered in this
letter. This bound complements existing CCB bounds and relates the
bottom Yukawa coupling to soft SUSY breaking parameters (and the
\(\mu\)-parameter of the superpotential) which is qualitatively
different from existing traditional CCB bounds. The bottom Yukawa
coupling itself depends nontrivially on the SUSY spectrum by virtue of
threshold corrections for large \(\tan\beta\). A similar bound was
found for the distinct direction in field space where all the
\(D\)-terms vanish. The corresponding unstable solutions are rather
metastable and very long-lived. Moreover, the comparison with quantum
corrected potentials shows that even the metastable configurations
tend to be stabilized by the loop contributions. This strengthens the
previous bound in the explicit non-\(D\)-flat directions which stems
from immensely short-lived configurations that persist in the presence
of quantum corrections and is therefore more severe. The limitation to
\(D\)-flat directions in the scalar potential as usually performed
probably misses additional potentially dangerous directions.

We constrained ourselves to cases with only one non-standard \vev,
accordingly the exclusions would change once more directions are taken
into account. In those cases, however, the definition of flat directions
suffers from ambiguities which makes the derivation of an analytic bound
similar to Eq.~\eqref{eq:analyticexcl} unclear. Similarly, the
constraints can be extended to non-vanishing stop and stau \vev{}s as
has been done for the left-right mixing of staus~\cite{Hisano:2010re}.

\section*{Acknowledgments}
This work was supported by the \emph{Karlsruhe School of Elementary
  Particle and Astroparticle Physics: Science and Technology (KSETA)}.
The author thanks U.~Nierste for useful discussions on the topic and
him and M.~Spinrath for reading and commenting on the manuscript.

\singlespacing
\bibliography{Bibliography}
\end{document}